\documentclass[apj]{emulateapj}
\usepackage{amsmath}                % American Mathematical Society package
\usepackage{amsfonts}               % American Mathematical Society fonts
\usepackage{amssymb}                % American Mathematical Society symbol
\usepackage{epsfig}                 % EPS figures
\usepackage[para,online,flushleft]{threeparttable}

\newcommand{\cm}{{~\rm cm}}
\newcommand{\km}{{~\rm km}}
\newcommand{\s}{{~\rm s}}

\newcommand{\erg}{{~\rm erg}}

\newcommand{\nar}{{~\rm New Astronomy Reviews}}

\newcommand{\pasa}{{~\rm Publications of the Astronomical Society of Australia}}

\begin{document}

%\title{The necessity of including magnetic fields in simulating core collapse supernovae}
\title{Reviving the stalled shock by jittering jets in core collapse supernovae: The key role of magnetic fields}

\author{Noam Soker\altaffilmark{1,2}}

\altaffiltext{1}{Department of Physics, Technion -- Israel Institute of Technology, Haifa
32000, Israel; soker@physics.technion.ac.il}
\altaffiltext{2}{Guangdong Technion Israel Institute of Technology, Shantou 515069, Guangdong Province, China}

\begin{abstract}
I find that an ingredient that was added in a recent study to facilitate the delayed neutrino explosion mechanism of core collapse supernovae, namely, large scale perturbations in the pre-collapse core, has a larger positive influence on the jittering jets explosion mechanism. By following the specific angular momentum of the accreted mass on to the newly born neutron star, I find that the accreted mass is likely to form intermittent accretion belts and disks, although they might lack axisymmetrical structure. These accretion belts and disks are likely to launch jets, but this can be simulated only if magnetic fields are included in the numerical code, as well as high numerical resolution that follows the rotation of the newly born neutron star and the shear in the accretion flow. I also discuss the possibility that the rotation of the pre-collapse core is important in increasing the shear in the accretion flow, hence the amplification of the magnetic fields.  
 I call for a \textit{paradigm shift from a neutrino-driven explosion mechanism of massive stars to a jet-driven explosion mechanism aided by neutrino heating. } Such a paradigm shift will bring the recognition that to simulate core collapse supernovae one must use magneto-hydrodynamical numerical codes. 
\end{abstract}

% ==========================================================
\section{Introduction}
\label{sec:intro}
% ==========================================================

The delayed neutrino mechanism is the most well simulated potential explosion mechanism of core collapse supernovae (CCSNe; e.g., \citealt{Bruennetal2016, Jankaetal2016, Muller2016, Burrowsetal2017}). However, this mechanism has problems (e.g., \citealt{Papishetal2015, Kushnir2015b}).
What I view as a more promising mechanism to explode CCSNe is the contesting jittering jets explosion mechanism \citep{PapishSoker2011}, or more generally the jet feedback mechanism (JFM; for a review see \citealt{Soker2016Rev}).  Neutrino heating does play a significant role in the JFM by keeping the outflowing gas hot, and by that adding energy to the bipolar outflow. 

Polarizations of some CCSNe and morphological features in some  supernova remnants (SNRs) indicate that jets play a role in many, or even most, CCSNe (e.g., \citealt{Wangetal2001, Maundetal2007, Lopezetal2011, Milisavljevic2013, Gonzalezetal2014, Lopezetal2014, Marguttietal2014, FesenMilisavljevic2016, Inserraetal2016, Mauerhanetal2017, GrichenerSoker2017, BearSoker2017, Bearetal2017, BearSoker2018, LopezFesen2018}).  \cite{BearSoker2018} argue that the morphology of SN~1987A, that is turning now to a SNR, is compatible with varying directions of the jets, as expected in the jittering jets explosion mechanism.
Numerical simulations and analytical studies, on the other hand, were limited mainly to the cases of rapidly rotating pre-collapse core, where the jets that are launched maintain a constant direction (e.g., \citealt{Khokhlovetal1999, Aloyetal2000, Hoflich2001, MacFadyen2001, Zhangetal2003, Woosley2005, Obergaulingeretal2006, Burrows2007, Couch2011, Nagakuraetal2011, TakiwakiKotake2011, Lazzati2012, Maedaetal2012, LopezCamaraetal2013, Mostaetal2014, LopezCamaraetal2014, Itoetal2015, 
Nishimura2015, BrombergTchekhovskoy2016, LopezCamaraetal2016, Nishimuraetal2017, Gilkis2018, Fengetal2018}).
Because pre-collapse rapid rotation requires in most cases that a stellar binary companion spins-up the core following a common envelope phase (at least in metal-rich stars), this case is not common. Traditionally, simulations of jets in CCSNe consider jet-driven explosion as rare.

Following the difficulties of the delayed neutrino mechanism, some new ingredients have been introduced to the explosion simulations, such as convection-driven perturbations (or turbulence) in the pre-collapse core (e.g., \citealt{CouchOtt2013, Mulleretal2017}).
The convection that introduces velocity fluctuation in the pre-collapse core can lead to angular momentum fluctuations of the mass accreted on to the newly born neutron star \citep{GilkisSoker2014, GilkisSoker2015, GilkisSoker2016}.
The point is that the pre-collpase convection that was argued to help the delayed neutrino explosion mechanism can have even more crucial roles in facilitating the jittering jets explosion mechanism. Indeed, in earlier papers (e.g., \citealt{GilkisSoker2015}) we argue that the convection (turbulence) that was introduced in simulations (e.g., \citealt{CouchOtt2013, CouchOtt2015, MuellerJanka2015}) is sufficient to form intermittent accretion disks/belts that can launch jets.

Motivated by the very important role of pre-collapse convection in the two  contesting explosion mechanisms, I continue the analysis of the effects of introducing perturbations from pre-collapse core convection. In a recent paper \cite{Mulleretal2017} introduce large-scale perturbations and simulate the outcome. They obtain an explosion within the frame of the delayed neutrino mechanism. However, the final baryonic mass of the neutron star was quite high, $1.85 M_\odot$. After reviewing some basics of the jittering jets explosion mechanism (section \ref{sec:jittering}), I analyze the specific angular momentum of the accreted gas in their simulations (section \ref{sec:AM}).  I discuss the energy budget in section \ref{sec:Energy}.  In my summary (section \ref{sec:summary}) I argue that the large specific angular momentum of the accreted mass should play an important role in facilitating the formation of jets, but this can be studied only if magnetic fields are included in the simulations.

% ==========================================================
\section{Angular momentum in the jittering jets explosion mechanism}
\label{sec:jittering}
% ==========================================================

The sources of the stochastic angular momentum of the mass that is accreted on to the newly born neutron star in the jittering jets explosion mechanism are velocity fluctuations in the pre-collapse core
\citep{GilkisSoker2014} that are amplified by instabilities in the post-shock region above the neutron star \citep{Papishetal2016}.
These instabilities include the spiral modes of the standing accretion shock instability (SASI; for studies of these modes, see, e.g., \citealt{BlondinMezzacappa2007, Rantsiouetal2011, Iwakamietal2014, Kurodaetal2014, Fernandez2015, Kazeronietal2017}), and turbulence driven by neutrino heating (e.g., \citealt{Kazeronietal2018}).
In the simulation of \cite{Mulleretal2017} that I analyze in section \ref{sec:AM} the seeds of the perturbations are set in the pre-collapse core, but the angular momentum fluctuations are largely amplified by the deformed shock and by the neutrino-heated bubbles that push the downflows around (M\"uller, B., private communication).

The stochastic specific angular momentum might lead to the formation of intermittent accretion disks or accretion belts. 
In cases where the specific angular momentum is large, but not large enough to allow the formation of a thin disk, e.g., the specific angular momentum is sub-Keplerian, an accretion belt is formed. The accretion belt does not extend much beyond the neutron star surface, and the accretion inflow is concentrated towards the equatorial plane, i.e., it avoids regions along the polar directions. In many cases the belt will not even possess an axi-symmetrical structure, e.g., it will be one sided. 

That last point is important. The amplification of magnetic fields can be done locally by both radial and tangential shear. So although a full belt is not necessarily formed (e.g., \citealt{Muller2015}) strong-sheared zones exist in the vicinity of the newly born neutron star. I speculate (based on \citealt{SchreierSoker2016}) that these local vortices can amplify magnetic fields that are then used to initiate the launching of jets. So from now on the reader should refer to the `belt' not as a fully developed rotating torus, but rather as locally rotationally sheared zones around the newly born neutron star. These local sheared zones suffer from large fluctuations in the velocity, both in magnitude and direction. 
Of course, my speculative claim that the belt can launch jets demand for simulations that include magnetic fields, very high numerical spacial resolution, and the full implementation of perturbations in the pre-collapse core. 

\cite{Papishetal2016} study the specific angular momentum of the accreted gas from the SASI results of \cite{Fernandez2010} and find the maximum specific angular momentum of the accreted gas to be $j \simeq 10^{15} \cm^2 \s^{-1}$. In an earlier paper \citep{Soker2017RAA} I examine the SASI results of \cite{Kazeronietal2017} and find the maximum specific angular momentum of the accreted gas to be $j \simeq 3 \times 10^{15} \cm^2 \s^{-1}$. The conclusion from these two studies is that fluctuations due to SASI can lead to the formation of intermittent accretion belts. However, these fluctuations are not as large as required to explain, for example, the morphology of the ejecta of SN~1987A \citep{BearSoker2018}.

In recent years some studies aiming to facilitate the explosion in the frame of the delayed neutrino mechanism introduced turbulence, resulting mainly from convection, to the pre-collapse core (e.g., \citealt{CouchOtt2013, CouchOtt2015, MuellerJanka2015, Mulleretal2017}). In turn, such a turbulence will ease the formation of intermittent accretion disks/belts around the newly born neutron star \citep{GilkisSoker2015}. In section \ref{sec:AM} I analyze the specific angular momentum of the accreted gas in the simulation of \cite{Mulleretal2017} where they introduce large-scale perturbations in a convective shell burning of their pre-collapse core model.  Before that I present here two effects of the angular momentum of the accreted mass, in forming low density regions along the two opposite polar directions and in amplifying magnetic fields.

Because of the centrifugal force, a mass element with a total specific angular momentum magnitude
\begin{equation}
j=\left( j^2_x+j^2_y+j^2_z \right)^{1/2} ,
\label{eq:j} 
\end{equation}
around an axis $\vec r$ cannot be accreted within an angle $\theta_a$ from the axis $\vec r$, given by \citep{Papishetal2016}
\begin{equation}
\begin{split}
\theta_a = &
\sin^{-1} \sqrt{\frac{j(t)}{j_\mathrm{Kep}}} =  \sin^{-1} \bigg[
0.31 \left( \frac{M_\mathrm{NS}}{1.6 M_\odot} \right)^{-1/4}
 \\
& \times
\left( \frac{R_\mathrm{NS}}{20 \km} \right)^{-1/4} 
\left( \frac{j(t)}{2 \times 10^{15} \cm^2 \s^{-1}} \right)^{1/2}
\bigg],
\end{split}
\label{eq:angle}
\end{equation}
where $R_\mathrm{NS}$ and $M_\mathrm{NS}$ are the radius and mass of the newly born neutron star, respectively, and $j_\mathrm{Kep} = \sqrt{GM_\mathrm{NS}R_\mathrm{NS}}$. 

One assumption that enters into equation (\ref{eq:angle}) is that the accreted gas has a uniform specific angular momentum. If it is not uniform, then gas with a specific angular momentum lower than the average value $j$ might flow along stream lines closer to $\vec r$, i.e., at an angle of $\theta<\theta_a$, while gas with a higher specific angular momentum forms a flatter accretion belt. The limiting angle $\theta_a$ serves as a representative behavior of the temporary accretion flow. In the simulation of \cite{Mulleretal2017} that I analyze in section \ref{sec:AM}, the accretion flow on to the neutron star is not axisymmetric, but rather is composed of downflows that hit the newly born neutron star obliquely. In that case  the angle $\theta_a$ does not represent a flow structure, but only represent the importance of the angular momentum of downflows. The shear that is induced by such a flow might play a crucial role in taping magnetic fields to eject gas from the neutron star vicinity. 

Two processes might further increase the value of the opening angle $\theta_a$, a dynamo in the accretion belt that amplifies the magnetic fields and neutrino heating \citep{Soker2017RAA}.
I give here an estimate of the magnetic field pressure $P_B$ as \cite{SchreierSoker2016} crudely estimate to be in a non-turbulence accretion belt,
\begin{equation}
\begin{split}
\frac {P_B}{\rho v^2_{\rm esc}} \approx &
\left( \frac{j}{j_\mathrm{Kep}} \right)^2 \approx 0.01  
\left( \frac{j_{\rm belt}}{2 \times 10^{15} \cm^2 \s^{-1}} \right)^{2} \\
& \times
\left( \frac{M_\mathrm{NS}}{1.4 M_\odot} \right)^{-1}
\left( \frac{R_\mathrm{NS}}{20 \km} \right)^{-1}, 
\end{split}
\label{eq:Pb}
\end{equation}
where $v_{\rm esc} \simeq 140,000 \km \s^{-1}$ is the escape velocity from the accretion belt and $\rho$ is the density in the belt.
This magnetic field that reaches about one to about ten percent  (see below) of the total energy per unit volume of the gas in the belt (gravitational + kinetic + thermal) might have a significant effect when considering the following points \citep{Soker2017RAA}.
(1) In the expected case of a turbulent accretion belt amplification will be higher. As well, as we see later, the value of $j=2 \times 10^{15} \cm^2 \s^{-1}$ is a lower value in the relevant time of accretion, and the typical value is larger by a factor of 2-10. Over all, for the case I study later 
the expected ratio is ${P_B}/{\rho v^2_{\rm esc}} \approx 0.1$. 
(2) It is sufficient that $\approx 5-10 \%$ of the $\approx 0.1 M_\odot$ that is accreted in the final mass accretion phase be launched in jets to obtain a typical CCSN energy.
(3) \cite{Endeveetal2012} find that outside the neutrinosphere the SASI can substantially increase the strength of the magnetic fields, implying relatively strong magnetic fields before they are amplified in the accretion belt or accretion disk. However, several simulations suggest that the amplification of the magnetic field by the SASI and other processes
 (e.g., \citealt{Obergaulingeretal2009, Obergaulingeretal2014, GuiletMuller2015, Rembiaszetal2016b}) might only moderately increase the magnetic field. I here consider the question of whether the amplification of the magnetic field is sufficient to explain jittering jets, and open question.

The above discussion suggests that when the accreted mass has a sub-Keplerian specific angular momentum that is non-negligible, magnetic fields cannot be ignored. I turn therefore to examine the recent simulations of \cite{Mulleretal2017}

% ==========================================================
\section{Specific angular momentum}
\label{sec:AM}
% ==========================================================

\begin{figure}
\includegraphics[width=0.52\textwidth]{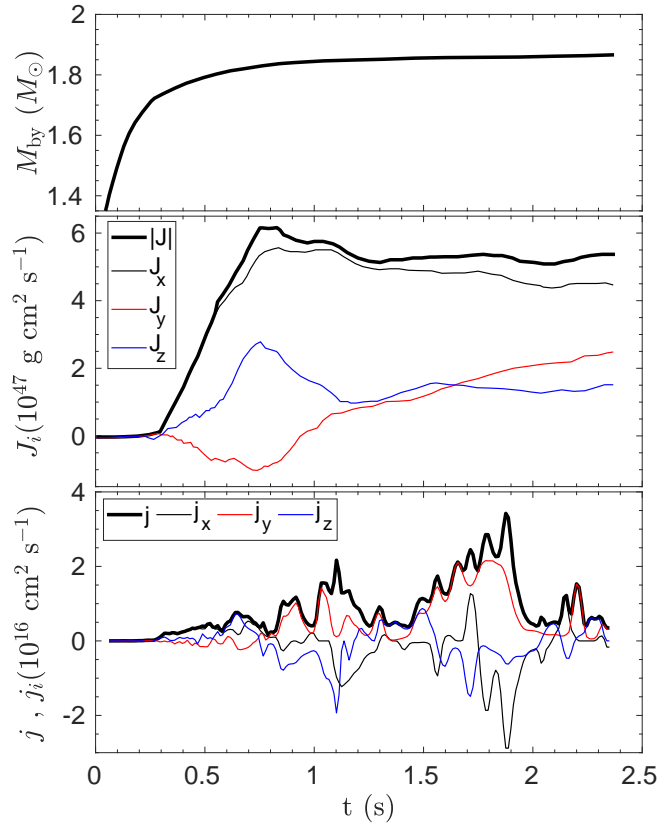}
\caption{The angular momentum of the accreted mass in a simulation performed by \cite{Mulleretal2017}. The two upper panels are from figure 20 there. The upper panel shows the baryonic mass of the neutron star and the second panel shows the components of the angular momentum of the neutron star and its total angular momentum. The lower panel is the specific angular momentum of the accreted gas that I calculate from the two upper panels by $j_i=dJ_i/dM_{\rm by}$, where $i=x,y,z$, while $j$ is calculated from equation (\ref{eq:j}). }
\label{fig:AccretedAM}
\end{figure}

I analyze the results of \cite{Mulleretal2017} concerning the angular momentum of the accreted mass. The upper two panels in Fig. \ref{fig:AccretedAM} are taken directly from their figure 20.
The upper panel shows the baryonic mass $M_{\rm by}$ of the neutron star as function of time, and the second panel is the angular momentum of the neutron star along the three Cartesian axes, $J_x$, $J_y$ and $J_z$, and the magnitude of the angular momentum J.
The lower panel is the specific angular momentum of the accreted gas that I calculate from the two upper graphs by $j_i=dJ_i/dM_{\rm by}$, where $i=x,y,z$, and then $j$ is calculated from equation (\ref{eq:j}). 
 Note that the angular momentum (second panel of Fig. \ref{fig:AccretedAM}) that is taken directly from \cite{Mulleretal2017} has many small fluctuations on time scales of few$\times0.01 \s$. When divided by the low accretion rate it gives the large fluctuations in the specific angular momentum that is seen in the lower panel.  
% FFFFFFFFFFFFFFFFFFFFFF
% FFFFFFFFFFFFFFFFFFFFFF

\begin{figure}
\includegraphics[width=0.52\textwidth]{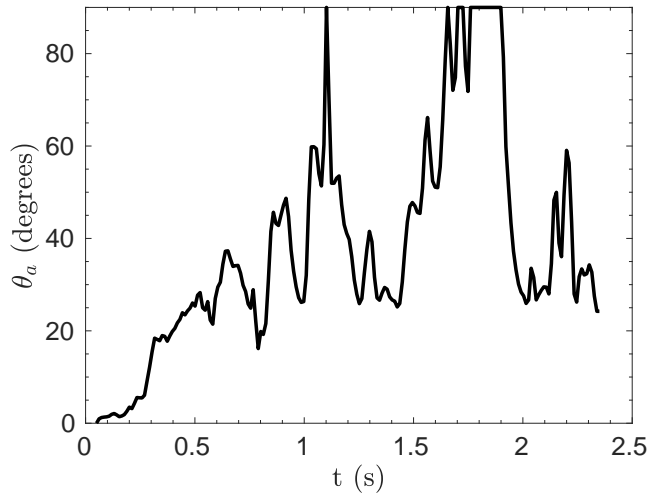}
\caption{The limiting angle from the angular momentum axis of the accretion flow according to equation (\ref{eq:angle}). The value of $j$ is from the thick line in the lower panel of Fig. \ref{fig:AccretedAM}.}
\label{fig:thetaA}
\end{figure}

In Fig. \ref{fig:thetaA} I present the limiting angle $\theta_a$ according to equation (\ref{eq:angle}) with the value of the specific angular momentum $j$ from the thick line in the lower panel of Fig. \ref{fig:AccretedAM}.
% FFFFFFFFFFFFFFFFFFFFFF
% FFFFFFFFFFFFFFFFFFFFFF

In what follows I scale the radius of the newly born neutron star with $R_{\rm NS}=20 \km$, and its gravitational mass with $M_{\rm NS}=1.6 M_\odot$, so that the gravitational potential on the surface is $\phi \equiv (G M_{\rm NS} / R_{\rm Ns} )^{1/2} = \left( 10^{10} \cm \s^{-1} \right)^2$.
If a fraction of about $\eta_j=0.05$ of the accreted mass is launched into the two jets at a terminal velocity which equals the escape speed, $v_{\rm esc} \simeq 140,000 \km \s^{-1}$, then for an accreted mass of $0.1 M_\odot$ the energy carried by the jets is $E_j \simeq 10^{51} \erg$.
Note that with this scaling the energy carried by the jets per unit accreted mass is $0.05 v^2_{\rm esc}$, which is compatible with the estimate ${P_B}/{\rho v^2_{\rm esc}} \approx 0.1$ from section \ref{sec:jittering}. 

From Figs. \ref{fig:AccretedAM} and \ref{fig:thetaA} I emphasize the following properties of the accretion flow.
  
(1) The typical coherence time of a fluctuation is $\approx 0.1 \s$. This equals about a hundred time the dynamical time on the surface of the newly born neutron star. This might suggest that the accretion flows has time to amplify the magnetic fields in the regions where the downflows hit the neutron star.   

(2) From $t=0.3 \s$ at least one component of the specific angular momentum has a magnitude of $j_i>2 \times 10^{15} \cm^2 \s^{-1}$, and the limiting angle has a value of $\theta_a \ga 20^\circ$ (Fig. \ref{fig:thetaA}). From that time on the amount of accreted mass is about $0.1 M_\odot$.
As stated above, such an accreted mass might launch jets with the desired explosion energy of $\approx 10^{51} \erg$.

(3) From about $t=0.5 \s$, when the final mass of $M_{\rm acc,f} \simeq 0.05M_\odot$ is accreted, the typical value of the specific angular momentum is  $j_i>4 \times 10^{15} \cm^2 \s^{-1}$, and Fig. \ref{fig:thetaA} shows that from that time on $\theta_a \ga 25^\circ$. 
This is quite a large angle. 

(4) There are time periods when at least one component has $j_i > 10^{16} \cm^2 \s^{-1}$, for which $\theta_a > 45^\circ$, as seen in Fig. \ref{fig:thetaA}. This can be considered as a thick accretion disk, even that it is formed from downflows and hence does not have an axi-symmetrical structure. There are times when $\theta_a=90^\circ$, namely, a thin accretion disk can form.  

(5) At about $t=1.55 \s$ to $1.85 \s$, i.e., for about $0.3 \s$, the specific angular momentum component $j_y$ has a large value of $j_y > 10^{16} \cm^2 \s^{-1}$.
The jets might maintain a more or less constant direction. The amount of accreted mass is small, $\approx 0.005 M_\odot$, but now a thick accretion disk is formed during this time period of $0.3 \s$. The energy in the jets that I expect during this activity phase is $\approx 10^{50} \erg$.
Jets that maintain a more or less constant direction and contain an energy that is $\approx 1-30 \%$ of the explosion energy can account for the presence of two opposite protrusions, called `ears', in some supernova remnants (\citealt{Bearetal2017, GrichenerSoker2017, YuFang2018}). It seems that the results of \cite{Mulleretal2017} might hint into such fixed-axis jets that the central object might launch at the end of the explosion process. 
  
  I note again that the accretion flow on to the newly born neutron star is not axisymmetric, but rather composed of downflows. These will form accretion belts that are not axisymmetric, e.g., one segment of a belt. But even such a flow requires a careful study that includes a high resolution numerical code that follows the rotation of the neutron star and includes magnetic fields.
 I can only crudely estimate the required numerical resolution. 
In principle the resolution should be high enough to resolve instabilities and to make numerical resistivity low. \cite{Rembiaszetal2016a} use a shearing box with cells' size down to about a meter. \cite{OConnorCouch2018} have cells' size of down to about 500 meter in their full 3D simulations. My estimate is that to have a first indication to the potential of launching jets, a full 3D simulation with cells' size of below about 50 meter is required.  
  
Despite this complicated accretion flow, I suggest that the behavior of the flow as revealed above is supportive of the jittering jets explosion mechanism. The real flow might be even more supportive. \cite{Mulleretal2017} introduce large-scale perturbations based on the 3D simulations of \cite{Mulleretal2016}. It is important to note that \cite{Mulleretal2016} simulated only the zone between $m=1.68 M_\odot$ and $m=4.07 M_\odot$, and this is where \cite{Mulleretal2017} introduced their perturbations. I expect that perturbations will be also presence in smaller mass coordinates (e.g., \citealt{Zilbermanetal2018}), so more mass and at earlier times than shown in Figs. \ref{fig:AccretedAM} is expected to be available to form accretion belts/disk.

% ==========================================================
\section{Energy considerations}
\label{sec:Energy}
% ==========================================================

 At this stage I can make only a crude estimate of the energy available for the explosion. First, the relevant rotational energy is the fluctuating one, not the one averaged over the entire accretion process and over the components of the angular momentum. 
As above, let me take the $M_{\rm acc,f}\simeq 0.05M_\odot$ that is accreted from about $t=0.5 \s$. The typical amplitude of the specific angular momentum (lower panel of Fig. \ref{fig:AccretedAM}) is $A_j \simeq 10^{16} \cm^2 \s^{-1}$. As above, I scale the radius of the newly born neutron star with $R_{\rm NS}=20 \km$. This implies that the typical rotation energy near the surface of the newly born neutron star is 
\begin{equation}
E_{\rm rotation} \simeq \frac{1}{2} M_{\rm acc,f} \left(\frac{A_j}{R_{\rm NS}} \right)^2 \approx 10^{51} \erg, 
\label{eq:Erot} 
\end{equation}
where in the last equality I substituted the typical values that I take from 
\cite{Mulleretal2017}.  

Equation (\ref{eq:Erot}) gives the crude total energy available from rotational shear. However, each jet of the many jittering jets will have less than 10 per cent of that energy. There are several to few tens of jets-launching episodes. In each episode the jets are not strong enough to penetrate the core, so they are choked inside the core and inflate high-pressure bubbles \citep{PapishSoker2014a, PapishSoker2014Planar}. These high-pressure bubbles merge to form a large high-pressure bubble that explodes the star by driving a shock wave through the outer core layers and the envelope. From this stage on the explosion is much like in the neutrino-driven explosion. For example, we expect the same nucleosynthesis outcomes. The general symmetry of the ejecta is different though. The last jets to be launched might leave a bipolar signature in the ejecta and in the supernova remnant (e.g., \citealt{GrichenerSoker2017}). But even these last jets do not break out from the ejecta, and hence the very high velocity jets cannot be observed, i.e., they do not leave the spectral signature of a very high speed gas in the ejecta. Only in extreme cases, like gamma ray bursts, the jets break out from the main ejecta. 

 There are other considerations that make the available energy larger than the value that equation (\ref{eq:Erot}) gives.
\newline
(1) \textit{Radial velocity.} In a Keplerian accretion disk the radial velocity of the accreted mass is negligible compared with the rotational velocity. Not here. Here the accretion on to the neutron star is not axisymmetric, but rather is composed of downflows that hit the newly born neutron star obliquely. The flow is neither spherically symmetric, implying that there is a shear from the radial velocity as well. Namely, part of the kinetic energy of the radial velocity component of the accreted mass is also available in principle for jet launching. One must include magnetic fields in the simulations to determine how much energy is available.   
\newline
(2) \textit{Neutrino energy.} When the jets penetrate through the stalled shock they actually locally revive the stalled shock. Near the stalled shock there is the gain region, namely, a neutrino-heat hot gas. This energy now becomes available to further push gas outward, but only along the direction of the jets. Along most other directions the stalled shock stays intact. Namely mass is continue to be accreted until the core is exploded. 
We have a jet-driven explosion aided by neutrino heating.
\newline
(3) \textit{Evolution of the jets' axis.} \cite{PapishSoker2014Planar} argued that because the jets eject core material along their propagation direction, latter accreted gas will flow-in mainly in direction perpendicular to the jets' axis. In present numerical simulations, despite the large amplitude of the angular momentum fluctuations, the average angular momentum is small because of the summation of many parcels of accreted gas. Each parcel has its angular momentum axis perpendicular to its inward radial direction. The accretion of gas with stochastic angular momentum from specific directions only increases therefore the average specific angular momentum. In the jittering jets explosion mechanism, after the first jets launching episode or two, the specific angular momentum increases, and jets might become more energetic even \citep{PapishSoker2014Planar}. 

Overall, it seems that there is a sufficient energy to explode massive stars with jets that are assisted by neutrino heating, although in most cases the high velocity jets cannot be observed directly. 

% ==========================================================
\section{Summary}
\label{sec:summary}
% ==========================================================

From the evolution of mass and angular momentum of the neutron star in the simulations of \cite{Mulleretal2017}, that I present in the two upper panels of Fig. \ref{fig:AccretedAM}, I calculated the specific angular momentum of the accreted mass as function of time (lower panel of Fig. \ref{fig:AccretedAM}). The value of the specific angular momentum is such that it might lead to the formation of an accretion belt or disk, although with no axisymmetrical structure because of the accretion flow of downflows. The possibility of intermittent disks and belts formation is shown by the value of the limiting angle $\theta_a$ that I defined in equation (\ref{eq:angle}) and presented in Fig. \ref{fig:thetaA}. 
 
The main result of this study is that the angular momentum of the accreted mass in the simulation of \cite{Mulleretal2017}, who introduced perturbations in the pre-collapse core, can lead to the formation of intermittent accreteion belts and disks around the newly born neutron star.

 The recent results of \cite{OConnorCouch2018} who find no explosion by neutrinos, but find very strong spiral-SASI modes with large angular momentum fluctuations, strengthen my claim. The implications of the results of \cite{OConnorCouch2018} to the jittering jets explosion mechanism are the subject of a forthcoming paper.  
  
 However, to derive bipolar outflows, namely jets, from this accretion flow one must include magnetic fields in the simulations, as well as high resolution that enables to follow the shear between the downflows themselves, and between the downflows and the rotating neutron star. The magnetic fields might be very strong. Firstly, the pre-collapse core is likely to amplify magnetic fields. 
\cite{Zilbermanetal2018} concluded from their study of the rotational shear in pre-collapsing cores that even slowly rotating pre-collapse cores might amplify magnetic fields in the core. 
Secondly, instabilities above the newly born neutron star can further amplify the magnetic field (e.g., \citealt{Endeveetal2012}). However, it is still an open question whether the amplification is large enough, as some simulations find limited magnetic field amplification (e.g., \citealt{Obergaulingeretal2009, GuiletMuller2015, Rembiaszetal2016b}).  Thirdly, the accretion disks (e.g., \citealt{Fujimotoetal2006}) and accretion belts (e.g., \citealt{SchreierSoker2016}; equation \ref{eq:Pb} above) substantially increase the magnetic field strength.
  
The study of \cite{Zilbermanetal2018} shows that pre-collapse cores possess some angular momentum. This raises the possibility that in addition to high resolution, future magneto-hydrodynamical numerical simulations will have to include rotation, even a small one, in the pre-collapse core. Namely, magneto-hydrodynamical effects in the collapse of a rotating core will lead to jets formation through very strong shear in the flow.    
  
Such magneto-hydrodynamical numerical simulations are possible. Some have already took the first steps towards full core collapse simulations with magnetic fields (e.g. \citealt{Masadaetal2015, Mostaetal2015, ObergaulingerAloy2017, Obergaulingeretal2018}), but these simulations do not include all necessary ingredients, e.g., pre-collapse perturbations. 
  
 I repeat again my call for a \textit{paradigm shift from a neutrino-driven explosion mechanism of CCSNe to a jet-driven explosion mechanism assisted by neutrino heating.} Most CCSNe are exploded by jittering jets,and then neutrino heating plays a significant role, but some are exploded by jets that maintain a fixed axis \citep{Soker2017RAA}, where neutrino heating plays a smaller role. In the present study this call is supported by the finding that an ingredient that was added to facilitate the delayed neutrino mechanism, namely, large scale perturbations in the pre-collapse core, has a larger positive influence on the jittering jets explosion mechanism. 
 
I thank Amit Kashi for his crucial help in obtaining the figures and for very helpful comments. I thank Avishai Gilkis for detail comments, and Bernhard M\"uller for some clarifications of the simulation. This research was supported by the E. and J. Bishop Research Fund at the Technion and by a grant from the Israel Science Foundation.

\end{document}